\documentclass[floatfix,useAMS,usenatbib]{mn2e}
 \usepackage{graphicx,epsfig}
\usepackage{multirow}
\usepackage{verbatim}
\usepackage[figuresright]{rotating}
\usepackage{txfonts}
\def\kms{km\,s$^{-1}$}
\def\msun{M$_{\odot}$}
\def\msunyr {M$_{\odot}~ \rm yr^{-1}$}

\title{Early X-ray emission from Type Ia supernovae originating 
from symbiotic progenitors or recurrent novae
}

\author[Georgios Dimitriadis, Alexandros Chiotellis and Jacco Vink ]{Georgios Dimitriadis$^{1,2}$\thanks{Email: G.Dimitriadis@soton.ac.uk}, Alexandros Chiotellis$^{1,3}$, Jacco Vink$^{1,4}$ \\ 
$^{1}$ Astronomical Institute ``Anton Pannekoek", University of Amsterdam, P.O. Box 94249, 1090 GE Amsterdam, The Netherlands\\
$^{2}$ School of Physics and Astronomy, University of Southampton, Southampton SO17 1BJ, UK\\
$^{3}$ Current address: National Observatory of Athens, Lofos Nymphon - Thissio, P.O. Box 20048, 11810, Athens, Greece\\
$^{4}$ GRAPPA,  University of Amsterdam, P.O. Box 94249, 1090 GE Amsterdam, The Netherlands
}

\date{Submitted 27 January 2014}
\pagerange{\pageref{firstpage}--\pageref{lastpage}} \pubyear{2014}

\begin{document}

\label{firstpage}

\maketitle

\begin{abstract}
One of the key observables for determining the progenitor nature of Type Ia supernovae is provided by their immediate circumstellar medium, which according
to several models should be shaped by the progenitor binary system.
So far, X-ray and radio observations indicate that the surroundings
are very tenuous, producing severe upper-limits on the mass loss from winds
of the progenitors.		
In this study, we perform numerical hydro-dynamical simulations of the interaction of the SN ejecta with circumstellar structures 
formed by possible mass outflows from the progenitor systems and we estimate numerically the expected numerical X-ray luminosity. We consider two kinds of circumstellar structures: a) A circumstellar medium formed by the donor star's stellar wind, in case of a symbiotic binary progenitor system; b) A circumstellar medium shaped by the interaction of the slow wind of the donor star with 
consecutive nova outbursts for the case of a symbiotic recurrent nova progenitor system. 
For the hydro-simulations we used well-known Type Ia supernova explosion models,
as well as an  approximation based on a power law model for the density
structure of the outer ejecta.
We confirm the strict upper limits on stellar wind mass loss,
provided by simplified interpretations of
X-ray upper limits of Type Ia supernovae. However, we show
that supernova explosions going off in the cavities created by repeated
nova explosions, provide a possible explanation for the lack
of X-ray emission from supernovae originating from symbiotic binaries.
Moreover, the velocity structure of circumstellar medium, shaped
by a series of nova explosion matches well with the Na absorption features
seen in absorption toward several Type Ia supernovae.
\end{abstract}

\begin{keywords}
hydrodynamics -- supernovae: general -- novae, cataclysmic variables

\end{keywords}

\section{Introduction}

Despite decades of research, the exact nature of the type Ia supernova (SN Ia)
progenitors remains unknown, making it one of the key unsolved problems in 
stellar evolution.
This lack of basic information regarding SNe Ia is quite remarkable,
given that 
SNe Ia are cosmological standard candles, which have provided 
key evidence that the expansion of the Universe is accelerating
\citep{perlmutter98,garnavich98}, and given the importance
of SNe Ia for the chemical evolution of the Universe, as they are a dominant
source of iron-group elements \citep{kauffmann}.

There is a consensus that SNe Ia are the thermonuclear explosions of
CO white dwarfs \citep[WDs,][]{bloom12}, but within this framework there are
two main scenarios for the progenitor binary systems that may lead to such
explosions:
1) the single-degenerate scenario (SD, \citealt{SD}) and 2)
the double-degenerate scenario (DD, \citealt{DD,Webbink1984}). 
In the SD scenario, a CO WD accretes hydrogen-rich or helium-rich material from a non-degenerate companion star. 
The companion could be either a main-sequence (MS) or a near main-sequence star in a close binary, or a star in the red giant branch (RG) or in the asymptotic giant branch (AGB) in a wider binary, called symbiotic systems. 
In the DD scenario, two CO WDs in a binary system are brought together by the loss
of angular momentum through the 
emission of gravitational radiation, after which they merge. Population synthesis studies of SN Ia suggest that the total population 
contains a mixture of these two possible scenarios, with the prompt channel mainly populated by SD progenitors, while the delayed channel mainly populated by DD 
progenitors \citep[e.g.][]{popsyn,claeys}.
	
	In order to test both scenarios, 
various studies have been done, which can be divided into two categories: 
a) directly observing the progenitor system or its left-overs; 
b) indirectly, by observing the effects of the
interaction of the SN with its surrounding. 
The direct methods include the detection of progenitor systems, 
either through the X-ray emission that accompanies the accretion process \citep{voss},
or by the identification of the donor star itself \citep{maoz}, either
in archival pre-explosion images, or by
searching for the surviving donor star in the centres of SN Ia 
supernova remnants \citep[SNRs,][]{ruizdonor,kerzendorf,schaefer}. 
While these methods could provide strong constraints on the 
properties of the progenitor system, the method can only be applied to nearby
galaxies.
These methods, therefore, provide limited statistics.
In contrast, the indirect methods are applicable to a larger number of SNe Ia, 
although the information they provide is related to the ambient medium in 
which the explosion occurs, rather than the progenitor system itself. 
	
	A promising indirect method is the study of the interaction of the SN blast wave with the ambient medium in which it expands. 
Considering the DD scenario, for which the components of the binary are two WDs, it is generally expected that  the ambient medium has been little affected
by the progenitor system. However, recently it has been suggested that circumstellar structures are also possible to be formed in the DD regime, either through successively nova explosions prior the the final merger  \citep{shen13}, or by the common envelope if the latter was ejected in the recent past of the SN Ia explosion  \citep{pakmor13, livio03}.

In contrast, for the SD scenario, substantial mass outflows are possible, 
 either from the wind of the secondary star, or from outflows
accompanying the accretion process, such as
WD accretion winds \citep{hachisu96} or nova explosions \citep{Sokoloski06}.
These various outflow mechanisms will modify the circumstellar medium (CSM)
substantially. The subsequent SN Ia will interact with this modified CSM and 
the result of the SN Ia/CSM interaction will be reflected in the 
optical, UV/X-ray and radio emission. 
				
The study of SN Ia/CSM interactions has, so far, provided 
contradicting results.
On the one hand several SN Ia show variable, blueshifted,
sodium absorption lines
\citep{patat,dilday,sternberg}, indicating that the SNe Ia were surrounded by
some sort of  dense expanding shell(s).
However, studies directed at identifying the interaction of
SN Ia blast waves with CSM modified by the stellar winds
of the companions show negative results.
Stellar winds result in a density profile that
scales as  $\rho\propto r^{-2}$, with  $r$ the distance from the wind source. This means that, 
in the early phase of the explosion, strong
emission is expected caused by the interaction of the blast wave
with the dense, inner regions of the wind.
This emission has not yet been detected in the optical \citep{mattila}, radio \citep{panagia,Chomiuk12} and X-rays (\citealt{immler}, \citealt{russel2012}),
providing strong constraints on the stellar wind mass-loss parameters.
For example, 
\citet{mattila} presented early time high-resolution and late time low-resolution optical spectra of the SN Ia, SN 2001el, and derived an upper limit for the mass-loss rate of the companion of $\dot{M}=9\times10^{-6}{\:}\rm{M_{\odot}{\:}yr^{-1}}$ and $\dot{M}=5\times10^{-5}{\:}\rm{M_{\odot}{\:}yr^{-1}}$ for wind velocities of $10{\:}\rm{km{\:}s^{-1}}$ and $50{\:}\rm{km{\:}s^{-1}}$ respectively. 
\citet{panagia} performed radio surveys of 27 SNe Ia with VLA and derived an upper limit of $\dot{M}=3\times10^{-8}{\:}\rm{M_{\odot}{\:}yr^{-1}}$. 
 The most constraining limit comes from the recent nearby SN Ia, 
SN 2011fe,\footnote{
While finishing this paper news broke out of a new nearby Type Ia supernova
SN2014J, in M82. This supernova will likely provide even more stringent
upper limits, or as we would hope a detection of radio and/or X-ray emission.
} for which the lack of radio emission in the early phase
implies an upper limit on the wind parameters of $\dot{M}=6\times10^{-10}{\:}\rm{M_{\odot}{\:}yr^{-1}} $  for a wind velocity of 
$u_w= 100$ \kms \citep{Chomiuk12}. 
\citet{russel2012} considered 53 SNe Ia observed with Swift X-ray Telescope, 
and their upper limit was 
$\dot{M}=1.1\times10^{-6}{\:}\rm{M_{\odot}{\:}yr^{-1}}$, based on a simple
model for the X-ray emission resulting from a blast-wave-wind interaction. 
Together 
these results exclude massive or evolved companions 
as donor stars, such as expected in the symbiotic system
scenario. But they do agree with small or main-sequence companions, 
or with double degenerate systems.
	
However, the circumstellar structure around the SN Ia could substantially deviate from a $\rho\thicksim r^{-2}$ density profile, because of other mass outflow mechanisms, emanating from the progenitor system, e.g. nova explosions. It has been shown that, in contrast with classical novae, where the WD mass is decreasing after every nova outburst, recurrent novae (RNe) may eventually
result in SN Ia explosions, as the WD is steadily increasing its mass 
\citep{rnpro}. These symbiotic RNe systems appear to be promising SN Ia progenitors, with the most famous example of potential
SN Ia progenitor being the nova RS Oph \citep{rsoph},
whereas an example of a supernova originating from such a system being
PTF 11kx \citep{dilday}, which shows evidence for surrounding nova shells. 
Moreover, \citet{woodvasey} have proposed that prior to the SN 2002ic explosion, a nova shell from the RN progenitor system created an evacuated region around the explosion centre. 

\begin{figure}  
\begin{center}
        \includegraphics[width=0.5\textwidth]{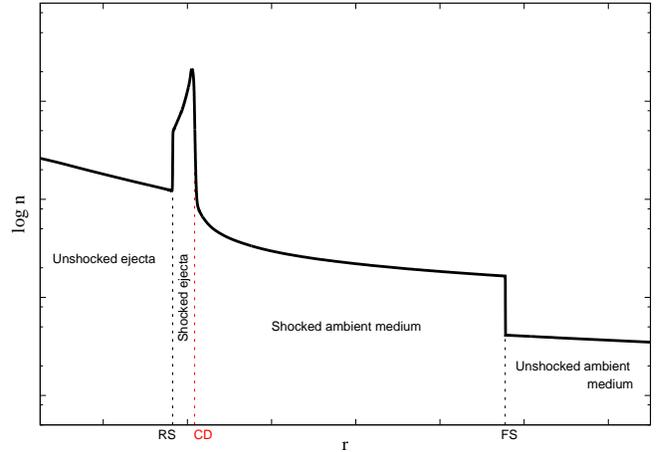}
         \caption{ The radial density profile of a SN evolving in a wind bubble. The `RS', `CD' and `FS' mark the positions of the reverse shock, contact discontinuity and forward shock respectively. }
  \label{fig:SNstruct}       
\end{center}
\end{figure}  

In this study, we report on our numerical hydrodynamic
studies of single-degenerate SN Ia explosions and their
interactions with the CSM shaped by either progenitor winds and/or
recurrent nova shells. 
In Sect. \ref{sec:windXray} we investigate the mechanism of the early X-ray emission for a SN Ia interacting with CSM formed by the donor star's stellar wind
 in a symbiotic binary progenitor system. We calculate the expected numerical luminosity of the X-rays and we re-evaluate previous studies on the specific energy band. 
In Sect. \ref{sec:SyRNe} we discuss the symbiotic recurrent nova system and we develop a model for the structure of the CSM shaped by consecutive nova outbursts. 
Moreover, we calculate the expected numerical luminosities of a SN Ia occurring in such a circumstellar structure. 
In Sect. \ref{sec:disc}, we discuss the implications of our study, verifying the X-ray upper limits in the case of the symbiotic progenitor system, and providing a physical explanation regarding the non-detection of the thermal X-rays in the case of a symbiotic recurrent nova progenitor system. 
In Sect. \ref{sec:sum} we summarise our results.

  \section{The early X-ray emission}\label{sec:windXray}

We model the X-ray radiation during the early SN phase ($\thicksim$150 days after the explosion) considering that the SN is expanding inside a wind bubble, shaped by the stellar wind of the donor star.
As such this study is a refinement of the model of \citet{immler}, who
used some approximations for the blast wave properties and the X-ray emission
resulting from the blast-wave/CSM interactions.
	
The general structure of a SN that expands in such an  environment is 
illustrated
 in Fig.~ \ref{fig:SNstruct}. From the centre outward we have: 
a) the freely expanding material ejected by the SN, 
b) the shocked ejecta shell, in which the ejected material is slowed down by the reverse shock, 
c) a contact discontinuity that separates the shocked ejecta material from the shocked ambient-medium material, d) the forward shock, which propagates supersonically into the undisturbed ambient medium, and e) the unshocked ambient medium.
	 
	  In this kind of configuration, X-rays are expected to be emitted from the region between the reverse and the forward shock, as there the 
plasma has been heated and compressed by the two shocks.
The X-ray luminosity, $L_x$, is 

\begin{equation}\label{eq:Lx}
	L_{x}=\int n_e n_i \Lambda_x(T,X)\mathrm{d}V ,
\end{equation}
 where $n_e$, $n_i$ are respectively the electron and ion  number density of the shocked plasma,  $\Lambda_x(T,X)$ is the cooling function in X-rays of the plasma with composition $X$ and temperature $T$ , 
and $V$ is the volume of the emitting material. 
	 
	 In the stellar wind scenario, the density of the unshocked wind is given by $\rho_{w}=\dot{M}/4\pi u_{w}r^{2}$, where $\dot{M}$ is the mass-loss rate of the donor star, $u_{w}$ the terminal velocity of the stellar wind and $r$ is the distance from the mass losing object. \citet{immler} studied the early X-ray properties of eight well observed SNe Ia aiming to find traces of such a SN - stellar wind interaction.  To correlate the observed X-ray emission with the stellar wind properties of the donor star the authors made the following assumptions. 
They considered that the blast wave is moving with a velocity of 
$u_{s}=40000{\:}\rm{km{\:}s^{-1}}$ \citep{chevalier1}, 
shock-heating the stellar wind to a constant temperature of $T=10^{9}{\:}\rm{K}$. For the 0.3-2 keV X-ray emission they
assumed a constant cooling function of
$\Lambda_{0.3-2{\:}\rm{keV}}(T)=3\times10^{-23}{\:}\rm{erg{\:}cm^{3} s^{-1}}$ \citep{fransson}.   
Substituting this in Eq. \ref{eq:Lx}, they obtained for 
the X-ray emission from the shocked stellar wind shell

\begin{equation}
	L_{x}=\frac{1}{\pi m^{2}}\Lambda(T) \left( \frac{\dot{M}}{u_{w}}\right)^{2} \frac{1}{u_{s}t},
\end{equation}  
where $m$ is the mean mass per particle ($2.1\times10^{-24} \rm{g}$ for a H+He plasma).
	
Finally, they considered that the X-ray luminosity of the shocked ejecta is 30 times higher than that of the shocked-wind shell \citep{chevalier1}.  
Using the possible detection of early X-ray
emission from SN 2005ke in the 0.3 - 2 keV band, 
they estimated the mass-loss rate of the progenitor's system to be $\dot{M}=3\times10^{-6}{\:} \rm{M_{\odot}yr^{-1}}$, assuming
a terminal velocity of the stellar wind equal to $u_{w}=10{\:}\rm{km{\:}s^{-1}}$. None of the other seven SNe Ia was detected in X-rays, and thus, they placed this mass loss estimation as the upper limit that a donor star can have. 

\subsection{Hydrodynamical Simulations}

In order to refine the estimates of \citet{immler} and \citet{russel2012}, 
we performed numerical simulations, using the hydrodynamic code AMRVAC \citep{amrvac}. AMRVAC solves the Euler equations in the conservative form, using an adaptive mesh refinement strategy, in order to refine the grid at specific positions, where large gradients in density and energy appear.

\begin{figure}
\begin{center}
        \includegraphics[width=0.5\textwidth]{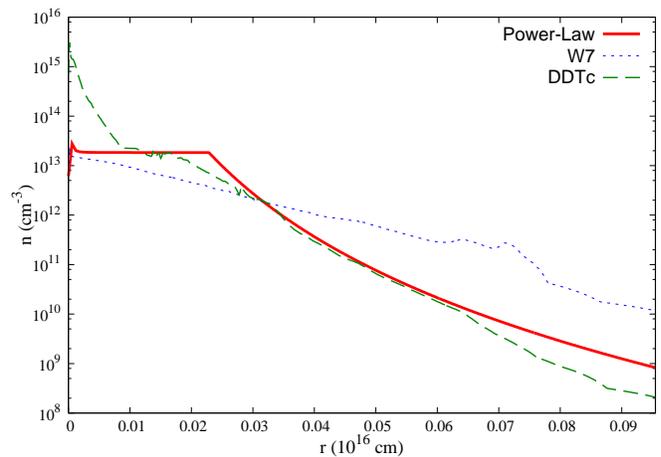}
         \caption{Initial ejecta density distribution of the three studied models. }
         \label{fig: SNejectamodels}
\end{center}         
\end{figure}

\begin{figure}
\begin{center}
        \includegraphics[width=0.5\textwidth]{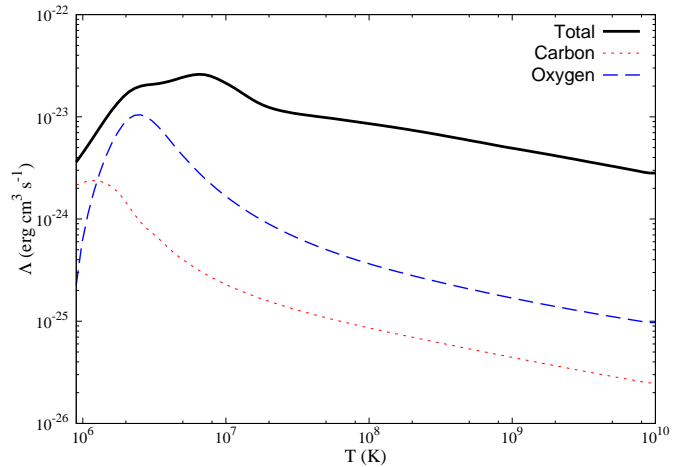}
         \caption{Cooling curves of a plasma of solar composition as a function of temperature, for the energy band of $0.3 - 2{\:}\rm{keV}$. The `total'  cooling curve corresponds to the cooling function  calculated for plasma with solar metallicity. For specific elements, only carbon and oxygen are depicted, since these elements dominate the chemical composition of the outer parts of the WD ejecta.}
         \label{fig:cooling}
\end{center}
\end{figure}
	
	We performed calculations on an one dimensional (1D) grid, with a radial span of $1\times10^{17} \rm{cm}$, assuming spherical symmetry. On the base level we used 540 grid cells and we allowed for seven refinement levels, at which of each the resolution is doubled. As a result, the maximum effective resolution becomes $3\times10^{12} \rm{cm}$.
	
\begin{figure*}
        \includegraphics[width=\textwidth]{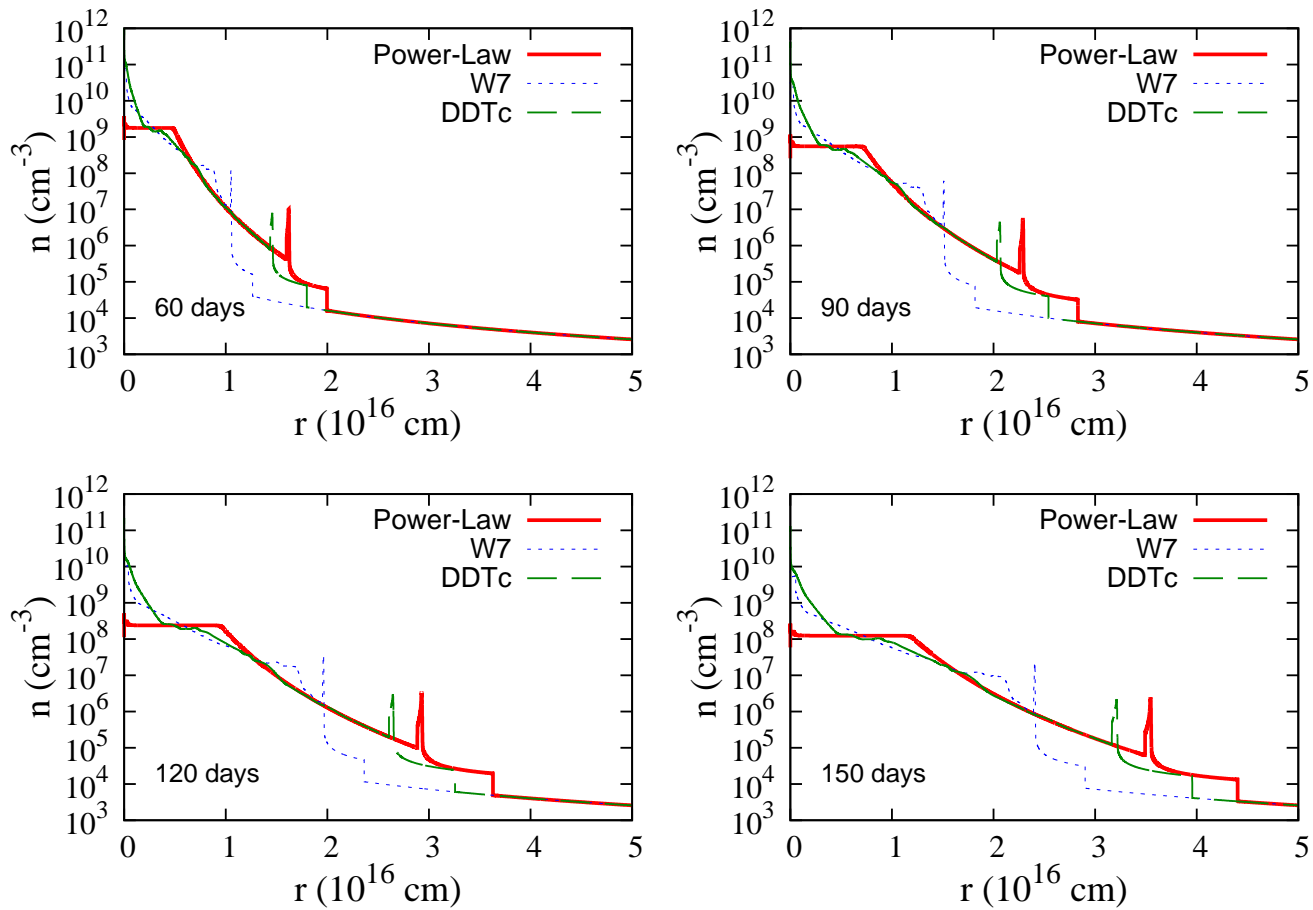}
         \caption{The SN density structure of the three studied models at four different time snapshots (60, 90, 120 and 150 days).}
         \label{SNmodels}
\end{figure*}
	 
	We filled the grid with a $\rho\thicksim r^{-2}$ density profile, representing the cold stellar wind bubble of the donor star $(T_{wind}=10^{3}{\:}\rm{K})$. We performed our calculations, using the wind parameters that \citet{immler} have derived, so the wind bubble  density is described as $\rho=\dot{M}/(4\pi u_{w}r^{2})$ with $\dot{M}=3\times10^{-6}{\:}\rm{M_{\odot}{\:}yr^{-1}}$ and $u_{w}=10{\:}\rm{km{\:}s^{-1}}$. At the inner boundary of the grid, we introduced  the SN ejecta. For our modelling we use three different SN Ia ejecta profiles.   Firstly, we used a power-law density profile with index $n=7$, with $M_{ej}=1.38\rm{M_{\odot}}$ and $E_{ej}=1.3\times10^{51} \rm{erg}$. The specific density distribution $\rho \propto r^{-7}$ is widely used for SN Ia studies \citep{chev}. In addition, we used two more ejecta density distributions, derived from two SN Ia explosion models: 
the W7 model \citep{w7}, a carbon deflagration model; and the DDTc model \citep{ddtc}, a delayed detonation model. 
The initial ejecta density distributions are presented in Fig. \ref{fig: SNejectamodels}. We advanced the supernova calculations for a time  interval corresponding
to 150 days. AMRVAC provides all the thermodynamical properties of the resulted SN structure, for selected snapshots in time. Moreover, we implemented a routine in order to locate the position of the reverse shock, the forward shock and the contact discontinuity. This was achieved by taking into account the Rankine-Hugoniot jump conditions and by introducing so-called composition tracers, in order to distinguish, as the system advances, the ejecta from the circumstellar material. The outcome of the specific routine is the positions of the shocks and the contact discontinuity as a function of time.

Following the procedure developed by \citet{klara}, we constructed new cooling curves at the energy band where the observations of \citet{immler} have been conducted. We used the SPEX package, version 2.03.03 \citep{spex}, in order to calculate the emissivity of a plasma, assuming collisional ionisation equilibrium. We calculated spectra for a logarithmic grid of temperatures from $10^{5}$ to $10^{10}{\:}\rm{K}$, with a step of $\rm{log}T=0.05{\:}\rm{K}$, assuming solar abundances. The cooling curves are presented in Fig. \ref{fig:cooling}. The emissivity $\Lambda_x(T,X)$ is calculated by integrating the spectrum over the desired energy range of $0.3 - 2{\:}\rm{keV}$. We have also calculated the contribution of the different elements at the cooling function, in order to be able to construct cooling curves of the shocked ejecta for each studied model. 

	For the SN evolution modelling, we incorporated our cooling curves to
the hydrodynamical simulations. 
For the CSM, we assumed solar metallicity, whereas for the shocked ejecta
the chemical composition was directly obtained from the explosion models.
For the  power-law density-distribution model, 
we assumed a chemical composition of C-O at 50\% each\footnote{{ Since our study concerns the early phase of SN evolution we assumed that only the unprocessed, outer layers of the WD have been heated by the reverse shock.}}. The electron density is calculated by assuming that the plasma is fully ionised.  
		 
The X-ray luminosity is given by 	 

\[
	L_{X}=L_{rev}+L_{for} =
\]
\begin{equation}
~~~~~= \int_{rs}^{cd} n_{e}n_{i} \Lambda(T,X) 4\pi r^{2} \mathrm{d}r+\int_{cd}^{fs} n_{e}n_{i} \Lambda(T,X) 4\pi r^{2} \mathrm{d}r ,
\end{equation}
where $L_{rev}$ is the luminosity from the shocked ejecta, $L_{for}$ is the luminosity from the shocked CSM, and the subscripts $rs$, $cd$, $fs$ refer to the radius of the reverse shock, the contact discontinuity and the forward shock respectively.
		
\subsection{Results of the hydrodynamical models for SN-wind interactions}

Fig. \ref{SNmodels} shows four snapshots of the SN evolution at $t=60,90,120,150{\:}\rm{days}$.  As the SN evolves, the densities of the shocked ejecta and the shocked CSM decrease, while each model expands in the wind bubble with a different velocity. Note the quite different evolution of the W7 explosion model, compared with the power-law and the DDTc models.

\begin{figure}
\begin{center}
        \includegraphics[width=0.5\textwidth]{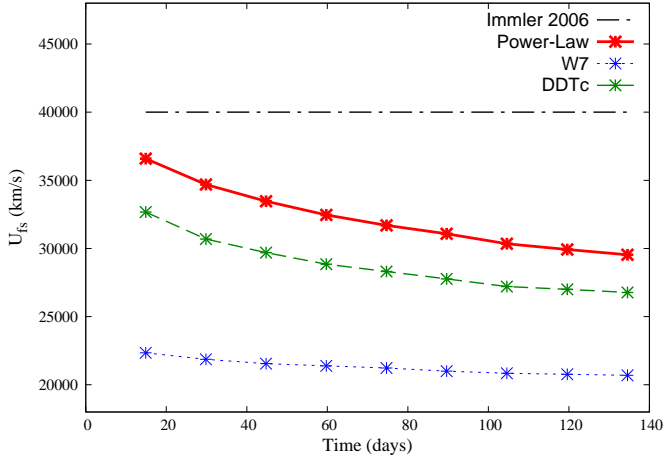}
         \caption{The time evolution of the  forward shock velocity for the three studied models. Comparison with the assumption of \citet{immler} (black horizontal line).  }
         \label{fig:SNvel} 
\end{center}
\end{figure}

Fig. \ref{fig:SNvel} shows the time evolution of the forward shock velocity. From this plot, it is clear that the assumption of a constant shock velocity of $40000{\:}\rm{km{\:}s^{-1}}$ by \citet{immler} overestimates the shock 
velocity as compared to our more detailed numerical calculations.
Moreover, in the numerical simulations
the shock velocity is not constant, but decreases as a function of time.

	Table \ref{table:SNvalues} lists the temperatures of the shocked CSM, the values of the (averaged) cooling function, the luminosity of the shocked CSM and the ratio $L_{rev}/L_{for}$ of both the numerical calculations and the values obtained by \citet{immler}.

\begin{table*}
 \centering
 \begin{tabular}{|c|c|c|c|c}
  \hline
   &$T(\rm{10^{10}~K})$&$\Lambda(T) (\rm{erg{\:}cm^{3} s^{-1}})$&$L_{for}(10^{37}{\:}\rm{erg{\:}s^{-1}})$&$L_{rev}/L_{for}$\\
    \hline
 	Immler&$0.1$&$3\times 10^{-23}$&$0.1-3$&$30$\\
	Power-Law&$1-3$&$3\times 10^{-24}$&$0.1-0.2$&$230-360$\\
         	W7&$0.7-1$&$6\times 10^{-24}$&$0.1-0.4$&$(3-8)\times 10^{3}$\\
         	DDTc&$1-2.5$&$3\times 10^{-24}$&$0.01-0.2$&$120-230$\\
 \hline
 \end{tabular}
  \caption{Re-evaluation of SN parameters derived by \citet{immler}.}
 \label{table:SNvalues}
\end{table*}

Based on our calculations, and depending on the explosion model we used, we find that the assumption for the temperature of the shocked CSM is an underestimation, since the numerical simulations indicate values that are an order of magnitude higher. For the cooling function, we find values $\thicksim 10^{-24}{\:}\rm{erg{\:}cm^{3} s^{-1}}$, which are an order of magnitude lower than \citet{immler}. The same deviation, an order of magnitude lower, is found for the luminosity of the shocked CSM. As for the ratio $L_{rev}/L_{for}$, we find that the assumption of $L_{rev}/L_{for}=30$ is an underestimation, as we find values $\geq 120$.

\begin{figure}
\begin{center}
        \includegraphics[width=0.5\textwidth]{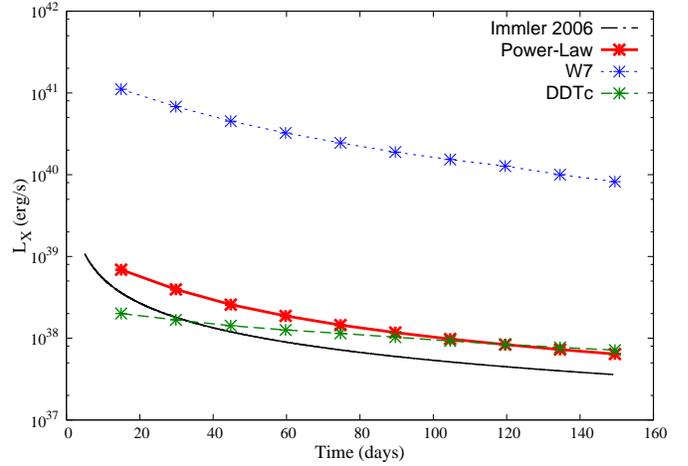}
         \caption{ The time evolution of the total SN X-ray luminosity for the three studied models in comparison with the estimation of \citet{immler}.}
         \label{fig:SNlum}
\end{center}
\end{figure}

Fig. \ref{fig:SNlum} shows the total X-ray luminosities  expected from the early phase of SN Ia, as a function of time. It is clear that, although we found deviations in our numerical results, regarding the assumptions made by \citet{immler}, eventually the total luminosity calculated in our models is similar for the DDTc and power law models and even higher for the W7 model. This similarity can be explained by taking into account the re-evaluation of the assumptions made by the authors. The total overestimation of the the cooling function and the luminosity of the shocked CSM is in the range of two orders of magnitude and is cancelled by the underestimation of $L_{rev}/L_{for}$. 
	 
\section{Symbiotic Nova systems}\label{sec:SyRNe}

Nova explosions are the result of thermonuclear runaway outbursts of the
WD's surface material, for those cases in which the accreted material
does not burn steadily, but builds up over time.
Theoretical studies show that the  nova outburst properties, such as
ejecta mass, ejecta composition and nova recurrence time, depend on 
the mass and temperature of the accreting WD, the composition
of the accreted material, and the mass accretion rate \citep{Yaron05,Townsley04}. 
Novae with more than one recorded outburst are called  recurrent novae (RNe). 
The RNe systems consist of a WD with mass close to the Chandrasekhar limit ($M_{WD}>1.3{\:}\rm{M_{\odot}}$) which accretes mass from its companion with a rate of ($\dot{M} \simeq 10^{-8} - 10^{-7} {\:}\rm{M_{\odot}{\:}yr^{-1}}$) \citep{livio92,Yaron05}.

The binaries that host recurrent novae  have been considered to be 
possible SNe Ia progenitors \citep{rnpro,woodvasey}
as in these systems the WD mass appears to be increasing. RNe can occur in both short period binaries (cataclysmic variables RNe, e.g. T Pyx) and in long period binaries (symbiotic RNe, e.g. RS Oph). 
In the latter case the donor star is a RG or an AGB. The resulting CSM around symbiotic RNe (SyRNe) is shaped by
the interaction of the 
continuous slow wind, emanating for the donor star, with the periodical, 
fast moving  nova ejecta. 

\citet{woodvasey} suggested that the lack of X-ray emission at the early SNe Ia phase is consistent with a  SyRNe progenitor  for which  the fast moving nova shells sweep up the dense wind material and form  a low density cavity around the binary system. Consequently, a low X-ray luminosity is expected from SN Ia resulting from these systems, as the blast wave of the SN Ia explosion, during the early phase, will interact with this low density
environment, rather than with 
the dense CSM formed by the donor's stellar wind. 

In this section, we investigate whether the suggestion of \cite{woodvasey} is feasible. For this purpose we simulate the CSM formed by  SyRNe and we calculate the  X-ray luminosity of the subsequent SNe Ia that are 
evolving in such SyRNe modified CSM. 

 \subsection{Hydrodynamical Simulations}

\subsubsection{Symbiotic recurrent novae}

 We performed calculations on an 1D grid, with a radial span of $2.5\times10^{18}{\:}\rm{cm}$, assuming spherical symmetry. At the base level we used 540 grid cells and we allowed for eight refinement levels, with each refinement doubling the resolution. As a result, the maximum effective resolution is $3\times10^{13}{\:}\rm{cm}$.  

Firstly, we simulated the CSM that is formed due to nova 
outbursts, interacting with the wind of the donor star. The final outcome depends on the wind properties ($\dot{M}, u_w$) and the RNe mass, energy and recurrence time. We used the wind parameters derived as upper limits by \citet{immler} ($\dot{M}=3\times10^{-6}{\:}\rm{M_{\odot}{\:}yr^{-1}}$, $u_{w}=10{\:}\rm{km{\:}s^{-1}}$) , 
in order to enable an easy comparison with a wind-only CSM. The wind is simulated by a continuous inflow at the inner boundary of the grid with density $\rho_{w}=\dot{M}/4\pi u_{w}r^{2}$ and momentum per unit mass $p= \rho \times u_w$. 

\begin{figure}
\begin{center}
        \includegraphics[width=0.5\textwidth]{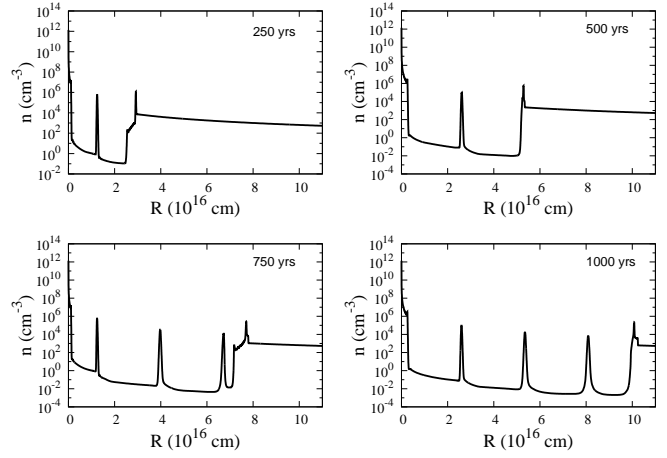}
         \caption{The time evolution of the recurrent nova cavity. In this simulation a recurrence time of 100~yr has been used.  }
         \label{fig:RNevol}
\end{center}
\end{figure}

In this grid we periodically  introduce the nova ejecta, at the inner boundary,   and we let it evolve. The nova ejecta mass is given by $M_{ej}=f\dot{M}_{acc}t_{rec}$, where $\dot{M}_{acc}$ the mass accretion rate, $t_{rec}$ the novae recurrence time and $f$ the fraction of the ejected mass over the accreted one. As for the case of RNe the WD increases its mass, $f<1$. Considering that the RNe recurrence time is of the order of 10 -100 yr, while the novae outburst occur for $\dot{M}_{acc} \sim 10^{-8} - 10^{-7}$ \msunyr, we get $M_{ej}= f \times (10^{-7} - 10^{-5})$ \msun.  The ejecta mass of the  2006 outburst of RS Oph was measured to be  $M_{ej} \simeq 10^{-7} - 10^{-6}$~\msun~ \citep{Sokoloski06,hachisu07}. In this work we use $M_{ej}=2\times 10^{-7}{\:}\rm{M_{\odot}}$. For the RNe explosion energy we adopt the value of \citet{novaparam}, estimated for the  the 2010 outburst of V407 Cyg: $E_{ej}= 2 \times 10^{44}$~erg. Finally, we describe the nova ejecta with a constant  density profile and a homologous velocity profile ($u_{ej}(R) \propto R$ ). Although this ejecta profile description is simplified, we claim that it does not affect the final outcome of our simulation as due to the low mass nova ejecta, the RN evolves to the Sedov - Taylor phase few days after the explosion \citep{moore12,Sokoloski06}. In this phase the evolution of the novae have lost any information about the initial properties of the ejecta. 

Figure \ref{fig:RNevol} depicts four snapshots of the nova cavity  evolution  in a SyRNe system. In this simulation we considered a nova recurrence time of 100~yr and we let the system advance for 1000 years.  The first nova explosion occurs and evolves in the dense wind bubble. Due to the low mass of the nova ejecta and the high density of the CSM, the nova passes to the Sedov - Taylor phase just one day after the explosion and in eleven days at the momentum driven phase. Due to the efficient cooling the shocked CSM collapses in a very thin shell, while the following nova ejecta has been homogenized as the nova reverse shock has reach the explosion centre.    A similar initial evolution is expected to be followed by the second nova explosion, as at the beginning the shock wave propagates in the dense wind that in the meanwhile has partially filled the cavity formed by the first nova. However, $\sim 1.5$ year later the nova forward shock penetrates the dense wind bubble and starts to propagate into the low density cavity formed by the first nova. Consequently, when the forward shock reaches the cavity it accelerates while the shocked wind shell expands in order to retain the pressure equilibrium.  Then the nova is moving with a constant velocity as the swept up mass of the nova cavity is negligible.  Inevitably, the second nova shell will collide with the first one. Due to this collision a reflected shock is formed which is propagating inwards - in the free expanding nova ejecta rest frame - till it will be swept up by the following nova. This process  is repeated for all the subsequent nova explosions. In the final snapshot of our simulation (bottom right of Fig. \ref{fig:RNevol}), six of the total nine nova shells have collided and merged with the edge of the cavity while the remaining three are still evolving in the cavity. 

\begin{figure*}
       \includegraphics[width=1.0\textwidth]{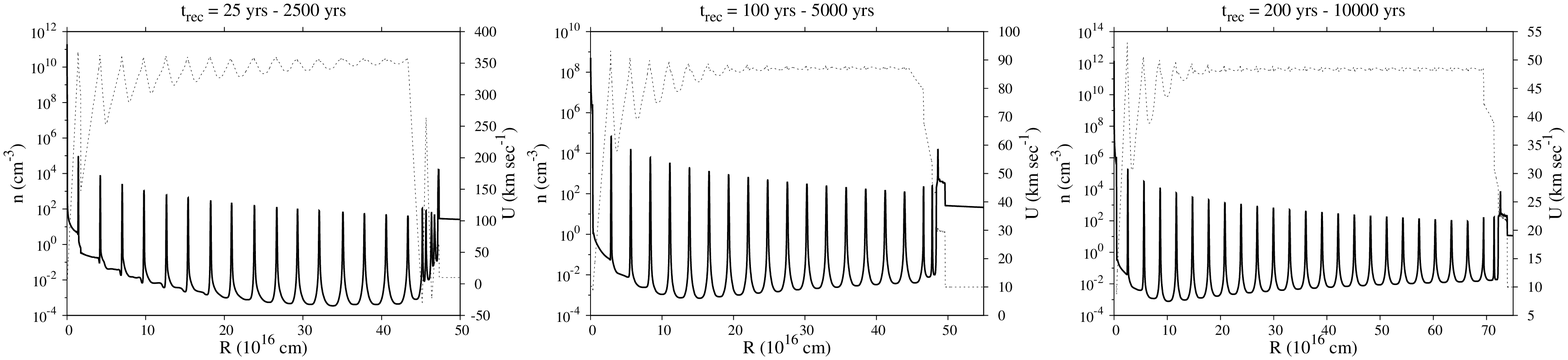}
        \includegraphics[width=1.0\textwidth]{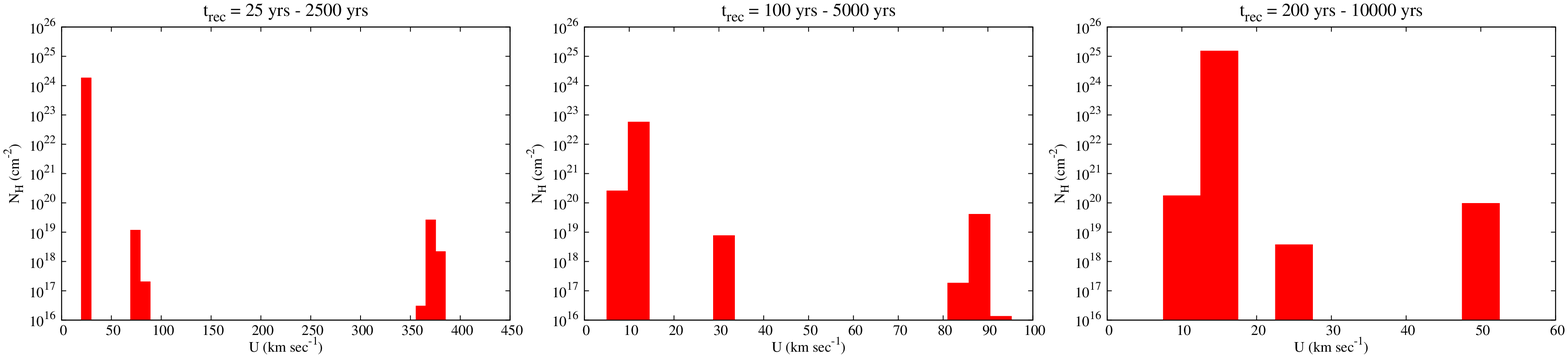}

        \caption{Upper row: The density (solid line) and the velocity (dotted line) of the CSM shaped by a SyReN system for three different nova recurrence time (25, 100 and 200 yr). Lower row: The histograms of the column density as a function of the plasma velocity for the three nova cavities depicted in the upper row. In this plots we take into account only the plasma component with $T \le 5 \times 10^3$~K (see text for details). }
\label{fig:RNcavities}
\end{figure*}

Fig. \ref{fig:RNcavities} (upper row) shows a more evolved nova cavity in a SyRNe system. To test the dependence of the CSM structure on the recurrence time, we  consider three cases:  $t_{rec}=$ 25, 100 and 200 yr. We  advanced the system for 2500, 5000 and 10000 years allowing for the formation and evolution of  99, 49 and 49 subsequent
novae, respectively. The general shape of the CSM is similar to that of Fig.~\ref{fig:RNevol}. Most of the novae have reached and collided with the edge of the nova cavity while the rest are still evolving with a constant velocity, into the low density medium, forming separate shells of shocked-wind material.   The densities in between the nova shells 
are tenuous - compared to the nova shells - in the order of $n\thicksim10^{-3} - 1 {\:}\rm{cm^{-3}}$. 

\begin{figure*}  
\begin{center}$
\begin{array}{cc}    
  \includegraphics[trim=0 0 0 10,clip=true,width=0.48\textwidth]{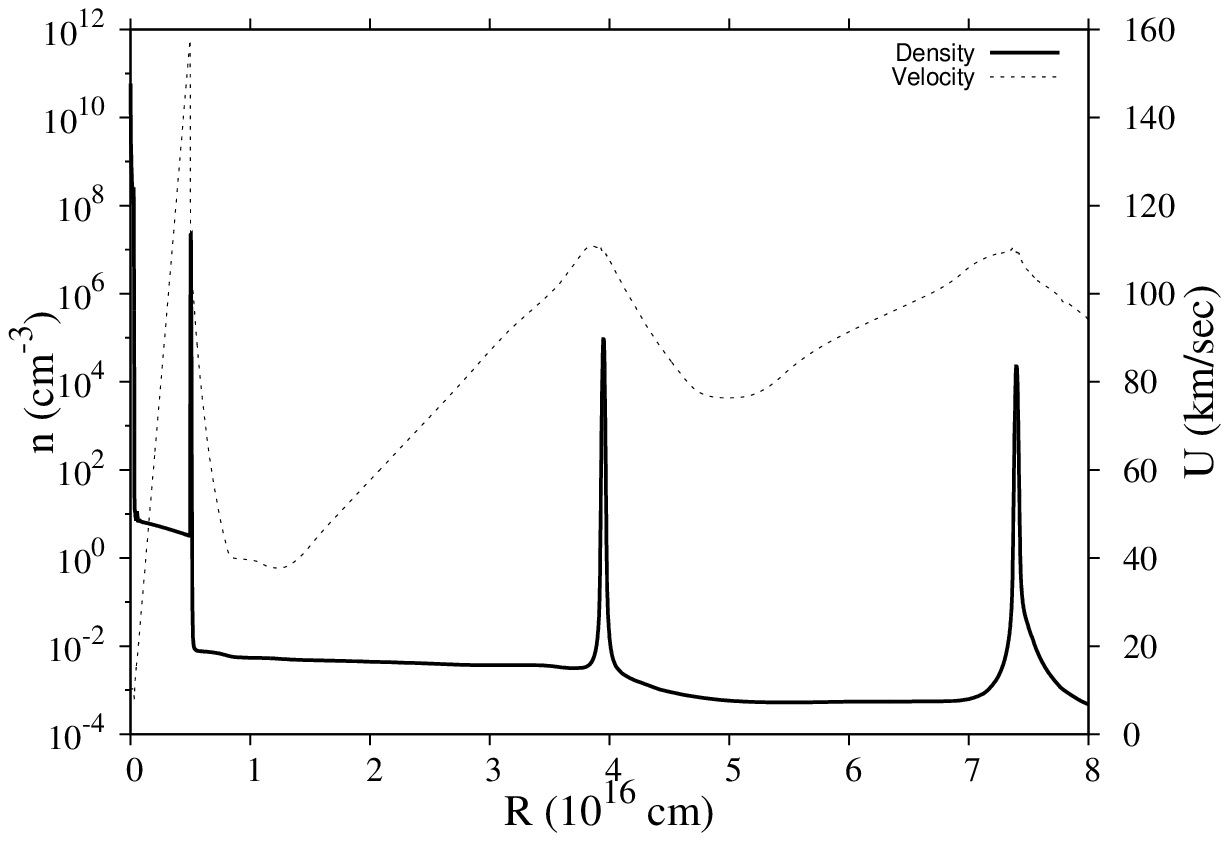} &
  \includegraphics[trim=0 0 0 10,clip=true,width=0.48\textwidth]{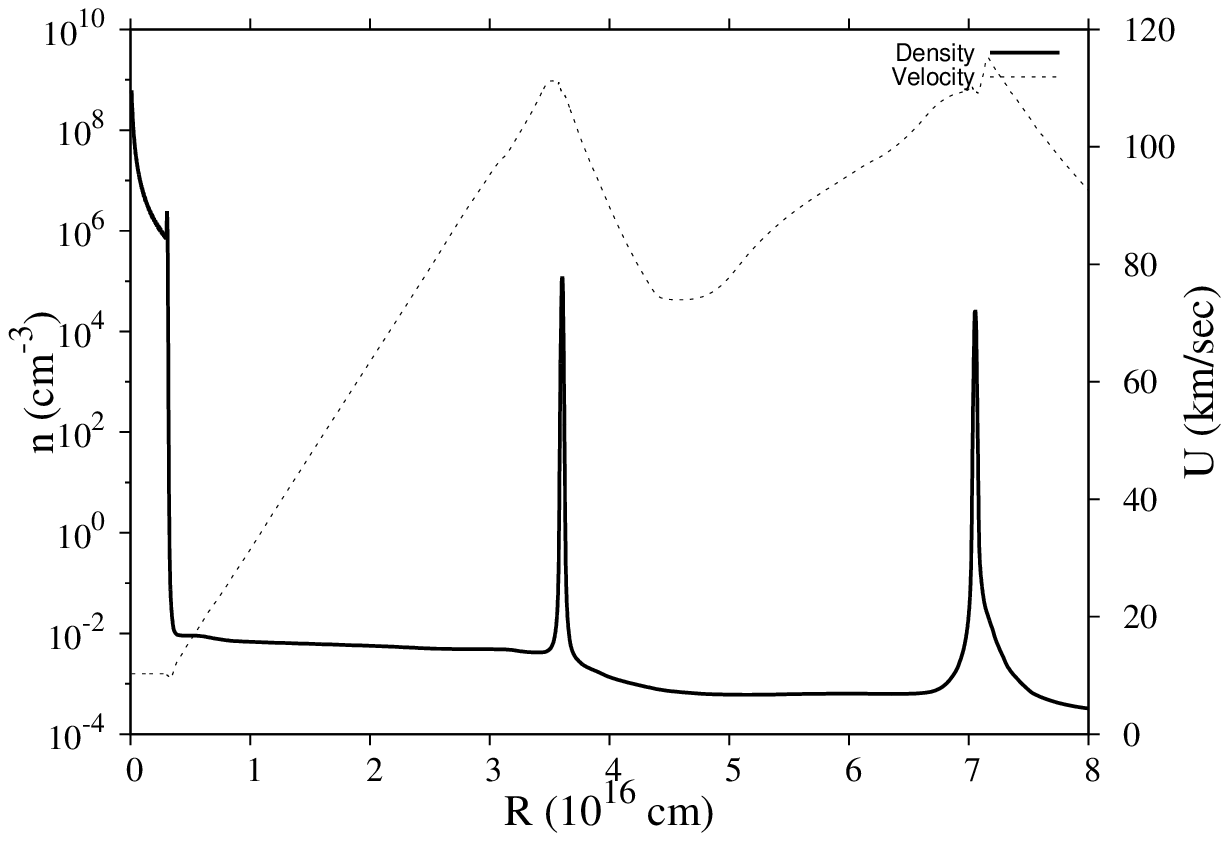} 

  \end{array}$
\end{center}
        \caption{The innermost region of CSM structure at the moment of the SN explosion.  For  Case 1  (left) the SN occurs right after the last nova explosion so at the inner boundary of the plot is no any wind structure. For Case 2 (right) the SN explosion occurs 100 yr after the last nova explosion so the wind has partially filled the nova cavity up to $4 \times 10^{15}$~cm. 
}
  \label{fig:RNcase12}

\end{figure*}

The nova recurrence time does not affect the general properties of the CSM. However, increasing the recurrence time, more wind material is entered into the cavity till the next outburst will occur. As a result, each nova sweeps up more wind material leading to a CSM characterized by more massive shells which are moving with lower velocities as compared to the cases with lower recurrence time. The cavity itself  evolves slower due to the high swept up mass and the more efficient radiation losses from the dense shell material. For these simulations, the cavities extend from $\thicksim4.5\times10^{17}{\:}\rm{cm}$ ($\thicksim0.15{\:}\rm{pc}$), up to $\thicksim7.5\times10^{17}{\:}\rm{cm}$  ($\thicksim0.25{\:}\rm{pc}$). The plasma velocity is ranging from $50{\:}\rm{km{\:}s^{-1}}$ up to $350{\:}\rm{km{\:}s^{-1}}$, depending on the recurrence time of the simulation.

\subsubsection{SN evolution in the nova cavity and its X-ray emission}

Having established the general structure of the CSM formed by a SyRN progenitor, we introduce the 
SN Ia ejecta in the inner boundary of the grid depicted in  Fig. \ref{fig:RNevol}. The evolution of the supernova blast-wave is similar for the three
recurrence times investigated. For illustrative purposes we limit ourselves 
here to the nova shells created by a nova with a recurrence time of $t_{rec}$= 100~yr.
For our calculations, we choose a small grid size of $2\times10^{17}{\:}\rm{cm}$ while we advanced the SyRN system up to one thousand years. Although  the SyRNe phase of Type Ia progenitors last up to $10^5$~yr revealing thousands to tens of thousands novae outbursts \citep{hachisukato08}, we show in the previous section that the CSM properties close to the explosion centre are not appreciable affected by the number of nova explosions and the time of the nova cavity evolution. Thus, without loss of generality, a small grid size allows us to resolve the inner structure of the supernovae with sufficient detail without being too computational expensive.  The maximum, effective resolution we get is $3\times10^{12}{\:}\rm{cm}$. 

The SN Ia ejecta density distribution is described by  a power law with index $n=7$, assuming a chemical composition of C-O at 50\% each.
As shown in Sect. \ref{sec:windXray}, this model is quite comparable with the DDTc model.
Similarly to the models described in section \ref{sec:windXray}, we calculated 
the expected X-ray luminosity of a SN Ia evolving in such a 
SyRNe modified CSM. The outer edge of the SN ejecta is at  $1 \times 10^{14}$ cm, corresponding to a starting time for the calculations of  $\sim 0.7$ days after the explosion.

We used two possible CSM structures, representing two extreme cases. For `Case 1' we advanced the SyRN system for 910 yr, and describes the case in which the nova explosion  occurred right before the SN Ia. In this configuration,  the donor star's stellar wind had too little time (10 years) to start filling the cavity and the innermost CSM is described by the low density nova ejecta . For `Case 2', the SyRNe's outflows  were evolving for 1000 yr. Therefore, the SN explosion occurred just before the subsequent nova explosion.  In this case, the wind has been filling the cavity for 100 years, so the SN Ia will first interact with material originating from the donor star's stellar wind. In Fig. \ref{fig:RNcase12},  the initial density distribution for the two cases is illustrated.  For Case 1, the SN will first interact with the small wind bubble of radius $3\times10^{14}\rm{cm}$, and then expand in the low dense nova cavity. For Case 2 the radius of the wind bubble is  $3\times10^{15}\rm{cm}$.

\begin{figure}
\begin{center}
        \includegraphics[width=0.5\textwidth]{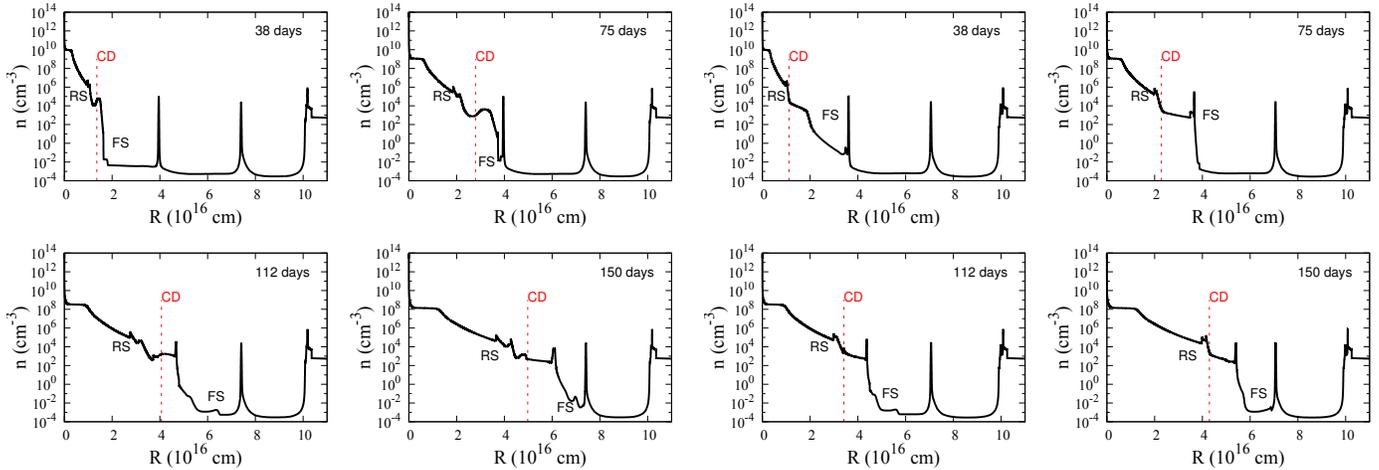}
         \caption{The density structure of the supernova evolving in the CSM of Case 1 (see text for details) at four different time snapshots. The symbols `FS', `CD', `RS', mark the positions of the SN forward shock, contact discontinuity and reverse shock respectively. The collision of the SN with the first two nova shells leads to the formation of a pair of transmitted - reflected shocks which are also visible in the figure. }
         \label{fig:SNcase1}
\end{center}
\end{figure}

\begin{figure}
\begin{center}
        \includegraphics[width=0.5\textwidth]{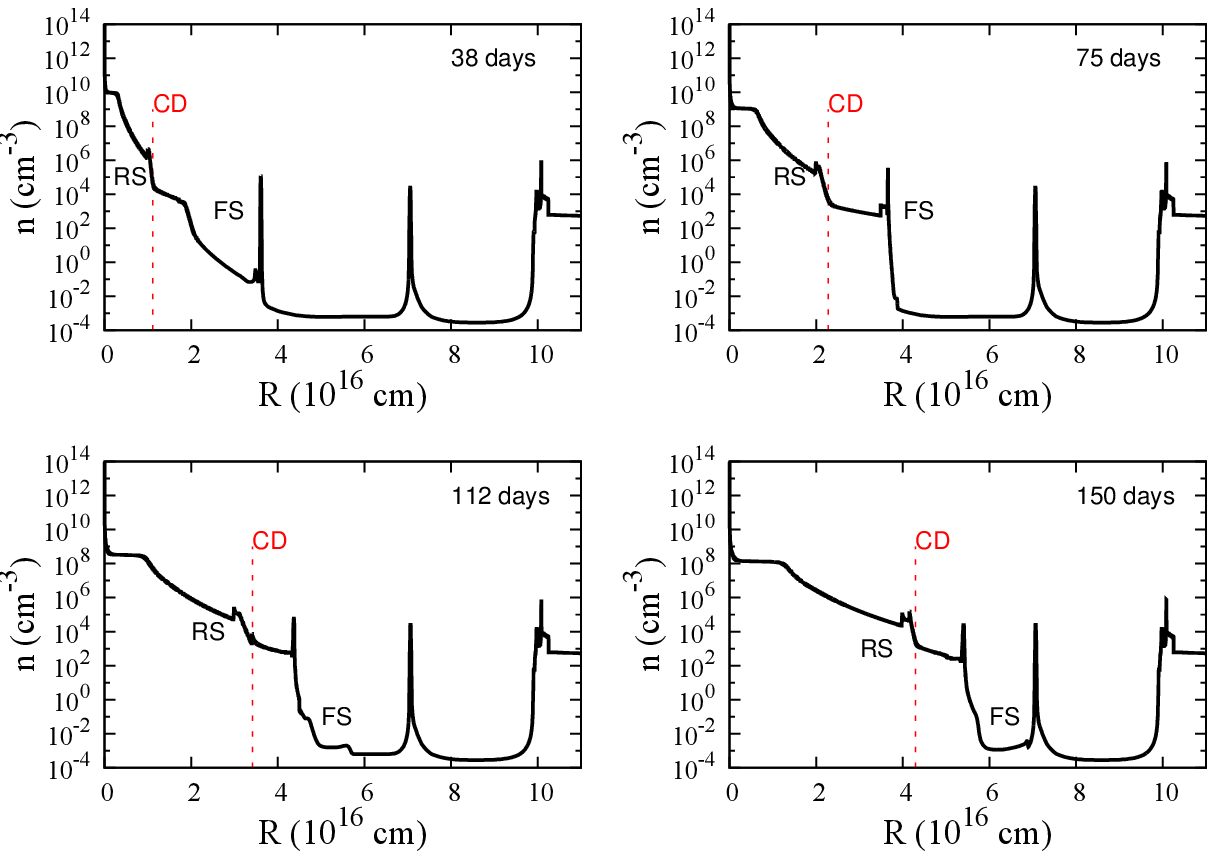}
         \caption{The same as Fig. \ref{fig:SNcase1} but for Case 2 (see text for details)}
         \label{fig:SNcase2}
\end{center}
\end{figure}

The evolution of the SN, for the Case 1 and Case 2, is shown in Fig. \ref{fig:SNcase1} and Fig. \ref{fig:SNcase2} respectively. For Case 1 (Fig. \ref{fig:SNcase1}), at the time of the SN Ia explosion, the first nova shell is located at a distance of  $r\simeq5\times 10^{15}{\:}\rm{cm}$, and the SN Ia starts interacting with this shell at $t\simeq20{\:}\rm{days}$. The collision of the SN Ia blast wave with the nova shell forms a pair of
reflected and transmitted shocks, which are propagating inward and outward respectively in the local plasma's rest frame.  At $t\simeq75{\:}\rm{days}$, the forward shock of the SN Ia reaches the second nova shell, located at distance $r\simeq4\times10^{16}{\:}\rm{cm}$. Our simulation ends at $t=150{\:}\rm{days}$, just before the SN Ia interacts with the third nova shell. 

For Case 2 (Fig. \ref{fig:SNcase2}), the SN Ia occurs just before the last nova explosion (at time $t=1000{\:}\rm{yr}$ of the simulation depicted in Fig. \ref{fig:RNevol}). 
The wind has filled the cavity for 100 years and its termination shock is located at $r\simeq3\times10^{15}{\:}\rm{cm}$. 
When the SN Ia occurs, it interacts with this wind bubble density profile, and it overcomes it within 10 days. The first  nova shell  is located at $r\simeq3.5\times10^{16}{\:}\rm{cm}$. 
It is  reached by the forward shock at $t\simeq35{\:}\rm{days}$. 
This collision creates a reflected shock that reaches the shocked ejecta shell at $t\simeq100{\:}\rm{days}$.

\begin{figure}
\begin{center}
        \includegraphics[width=0.5\textwidth]{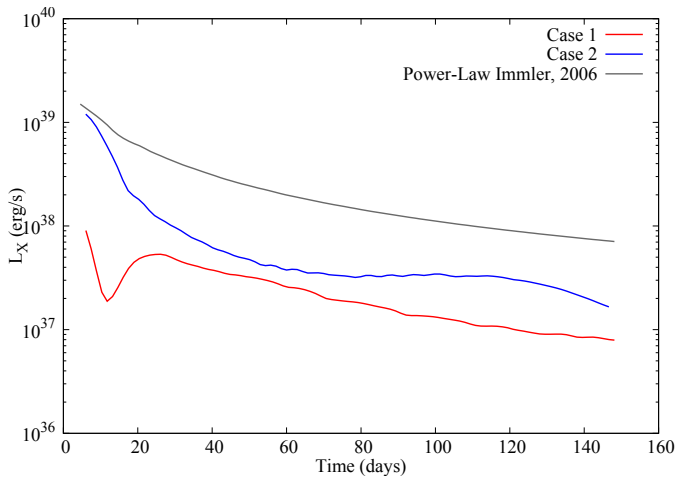}
         \caption{The time evolution of the total SN X-ray luminosity for a SN evolving in a nova cavity (Case 1 and Case 2) in comparison with the X-ray luminosity of a SN evolving in a wind bubble.   }
         \label{fig:SyRNelum}
\end{center}
\end{figure}

Fig. \ref{fig:SyRNelum} shows the total expected numerical X-ray luminosities $L_{X}$ as a function of time, for the Case 1 and 2 described above. In order to calculate more precisely the expected luminosities, we used a smaller time interval than the work that we performed in section \ref{sec:windXray}. The luminosities are calculated every 1.5 days, and an interpolation scheme  using of natural smoothing splines has been applied. The total expected numerical X-ray luminosity derived for a power law SN ejecta profile ejecta evolving in the stellar wind (Sect. \ref{sec:windXray}) is also presented for comparison.

For Case 1, the luminosity  at the start is decreasing, from $L_{X}\simeq10^{38}{\:}\rm{erg{\:}s^{-1}}$ down to $L_{X}\simeq2\times10^{37}{\:}\rm{erg{\:}s^{-1}}$. At time $t\simeq10{\:}\rm{days}$, the luminosity starts to increase, up to $L_{X}\simeq5\times10^{37}{\:}\rm{erg{\:}s^{-1}}$, reaching a maximum around $t\simeq25{\:}\rm{days}$, and then starts to decrease again, up to $L_{X}\simeq8\times10^{36}{\:}\rm{erg{\:}s^{-1}}$. This transition period at time $t\simeq10{\:}\rm{days}$ until $t\simeq25{\:}\rm{days}$ marks the time when the SN blast wave collides with the first nova shell and the reflected shock formed by this collision propagates inward heating the shocked ejecta and thus increasing its luminosity. 

For Case 2, the luminosity is initially similar to the wind-only model,
as the blast-wave interacts with the donor wind filling the nova cavity.
But the luminosity rapidly decreases as the blast wave reaches the end
of the wind, entering the nova cavity. As a result the
luminosity decreases from $L_{X}\simeq2\times10^{39}{\:}\rm{erg{\:}s^{-1}}$ 
down to $L_{X}\simeq2\times10^{37}{\:}\rm{erg{\:}s^{-1}}$. 

In general, the extracted luminosities of the SN Ia/SyRNe models reveal a more complex and non-monotonous function with time as compared to these of the SN Ia/wind models and they strongly depends on the time delay between the last nova and the SN Ia. However, for both Case 1 and Case 2  - which  represent the two extreme cases of this time delay - the SN Ia/SyRNe luminosity is at least one order of magnitude lower that this of the SN Ia/wind model. Thus,  we conclude that
the non-detection of X-rays following SN Ia explosions
can potentially be explained by a relatively low density environment  caused
by nova outbursts.     

\section{Discussion}\label{sec:disc}

In Sect.  \ref{sec:windXray}, we performed numerical calculations in order to investigate the thermal X-ray emission due to interaction of the SN Ia blast wave with CSM formed by the donor star's stellar wind.  We found that  even though the assumptions made by \citet{immler} deviate from the numerical results, eventually the total luminosities extracted from our model are similar, or higher, depending on the choice of the explosion model. Thus,  our work verifies the existing constraints on the mass loss rate of the donor star ( $\dot{M}<3\times10^{-6}{\:}\rm{M_{\odot}{\:}yr^{-1}}$) or even imposes stricter limits in the case of the W7 deflagration model.  Such a result excludes an evolved RG/AGB companion stars, in the SD regime, but  allows a main-sequence donor star, which stellar wind is much more tenuous ($\dot{M}\thicksim10^{-11}{\:}\rm{M_{\odot}{\:}yr^{-1}}$,  $u_{w}\thicksim1000{\:}\rm{km{\:}s^{-1}}$). Moreover, a double degenerate white dwarf system is also not excluded, as long as the ejection of the common envelope (CE) occurred well before the final merger of the two WDs, leaving a `clean' environment around the explosion centre.   

In Sect. \ref{sec:SyRNe}, we simulated the interaction of a SN Ia with the CSM shaped by the mass outflows of a SyRNe system.  The motivation for this simulation was the suggestion of \citet{woodvasey} that a nova explosion that occurred before the SN Ia, creates an evacuated region around the explosion centre and therefore can naturally explain the lack of X-ray emission at the early SN Ia phase. In addition, SyRNe systems have been proposed as viable candidates of SNe Ia progenitors  \citep{mikolajewska13,dilday,rsoph} and thus it is worthwhile to investigate if these systems are aligned with  the X-ray observations of SNe Ia. The outcome of our work is that nova outbursts are capable of creating a low dense ambient medium in which the subsequent SN Ia evolution is not detectable in X-rays. 
\\

\subsection{Similarities of SyRNe CSM with observed SNe Ia}

Although our intention was not to model any specific SN Ia,
and our approach of modelling a SN Ia explosion in a  SyRNe system is somewhat simplified 
(see \citealt{Mohamed13b}, for a detailed 3D simulation the SyRN RS Oph), 
the similarities between our SyRNe models and the recently observed properties of the CSM that appear to
surround a number of SNe Ia justifies a closer comparison.

The idea of an evacuated region around the explosion centre shaped by a nova explosion was first suggested to explain the density variations in the CSM of the   remarkable SN Ia  SN 2002ic \citep{woodvasey}.  
SN 2002ic showed no signs of CSM interaction during the first 5 - 20 days after the explosion 
(there are no observations prior to day 5). 
However, 22 days after the explosion, its brightness increased sharply revealing strong $\rm H_{\alpha}$-line
emission \citep{Hamuy03a}.   
A second brightening occurred  around 60 days after the explosion,
implying a second increase in the CSM density. Such a non-monotonous density distribution suggests 
the existence of periodic shells around the SN.  
\citet{woodvasey} showed that  a nova explosion in a symbiotic binary system
that took place $\sim 15$ years before the SN Ia forms an evacuated region with a radius of  $1.5  \times 10^{15}$~cm.  
Such a structure can explain brightness evolution during the first 22 days of SN 2002ic. 
Finally, they interpreted the secondary rise of the SN light curve (at 60 days) is the result of the collision of the 
SN blast wave with a second shell, formed by the previous nova explosion. Our SyRNe simulations reveal a similar
multiple-shell structure, while the first two shells are located to similar radii with these predicted by \citet{woodvasey}. 
For instance for the `Case 1 model' (see Fig.~\ref{fig:SNcase1}) 
the first nova shell is placed at a radius of $ 4  \times 10^{15}$~cm,  
and the second shell at $ 3.9  \times 10^{16}$~cm. The interaction of the SN with the first shell takes place for the
model $\sim 10$~days after the SN explosion, while the interaction with the second shell at 75 days (top right of Fig.~\ref{fig:SNcase1}).

Another recent SN Ia that reveals evidence of a multiple shell structure around the explosion centre is the PTF 2011kx \citep{dilday}. 
The time variability of the optical absorption lines detected in the spectra of this SN suggests
the existence of an inner shell at a radius of $1 \times 10^{16}$~cm with an expansion velocity of  
$\sim 100$ \kms, which was surrounded by a more distant, second shell moving at $\sim 65$ \kms. 
Two other SN Ia, SN 2006X \citep{patat} and SN 2007le \citep{simon09}, 
also showed a similar variability of optical absorption lines during the early SN phase, indicating the
presence a dense  CSM ($n \sim 10^{5}~ \rm cm^{-3}$, needed to satisfy the required recombination timescale)
at a distance of $10^{16} - 10^{17}$~cm from the explosion centre moving outwards with a velocity of 50 - 100 \kms.  

The results of our hydrodynamical simulations are in accordance with these observations. 
The histograms in Fig. \ref{fig:RNcavities}  show the column density distribution as a function of the plasma velocity for 
SyRNe CSM, depicted in the same figure. In these plots we took into account only the plasma 
with temperatures $T \le 5 \times 10^3$~K, as only for these temperatures neutral, absorbing material is expected. 
In the three cases, the column density distribution is characterized by three components. 
The low velocity component showing up around $u \simeq 10$ \kms, corresponds to the stellar wind that has partially filled 
the cavity, the medium velocity component corresponds  the plasma that has been accumulated at the edge of the cavity, 
and the high velocity component is formed by the first 2-3 shells inside the nova cavity, which are located
at radii of $(1 - 6)\times 10^{16}$~cm. 
The first component should not be observable in SN Ia, as by the time the SN is detected the blast wave has already
reached the end of the wind bubble. The second component is at a far distance from
the SN, and can therefore not contribute to the variability of the  absorption lines, as it will not be ionized by
the SN. Therefore, the component that is most relevant for the absorption lines in SN Ia is the third one. 
The velocity range of this component is $\thicksim50{\:}\rm{km{\:}s^{-1}}$ up to  $\thicksim350{\:}\rm{km{\:}s^{-1}}$ 
while  the corresponding column densities are in the range of $10^{18} - 10^{20} \rm cm^{-3}$. 
The velocity range is  in agreement with the observations.
As for the column density,
\citet{simon09} estimated that the Na column density around SN 2007le is around $2.5 \times 10^{12} \rm cm^{-2}$. 
Considering solar abundances for Na, this value corresponds to a total hydrogen column density toward the
the explosion centre of  $1.3 \times 10^{18} \rm cm^{-2}$, consistent with our simulations.

 \citet{shen13} attribute the absorption line
variability to shells formed by a sequence of nova explosions 
that potentially can  take  place prior to the final merger of the two white dwarfs, in the DD regime.
In their model, similar to ours, a shell of swept up ambient medium material is formed by the fast moving nova ejecta which at the moment of the SN Ia explosion has a radius and velocity similar to what is observed in SN 2006X, 2007le and PTF 2011kx.  However, in this model the nova ejecta are evolving in the interstellar medium, rather
than in a dense wind bubble. 
Consequently, in the absence of any kind of hydrodynamical instability or clumpiness, the cooling timescales of the nova shells are 
much higher than the dynamical ones. As a result, no efficient cooling takes place, and the temperatures of the shells 
are not consistent with the presence of neutral, absorbing material.
Moreover, this model can only produce one shell, as any subsequent nova is evolving from the start within 
the low density nova cavity, and without sweeping up any appreciable amount of mass, it will collide with the shell formed by previous novae. 

Recently \citet{soker13} suggested that the consecutive shells around PTF 2011kx can be also be explained by the violent merger scenario. According to this scenario, after a common envelope episode that takes place at the progenitor binary, the WD merges with the hot core that remained from the companion star. The multiple shell structure, in this case, is formed by the interaction of the wind of the AGB progenitor and the ejected mass of the common envelope. Hydrosimulations are needed in order a firm conclusion to be drawn about the validity of this model.  

 \section{Summary}\label{sec:sum}

The results of the present work are summarised as follows:
\begin{enumerate}
	\item 
We simulated the interaction of SNe Ia with circumstellar wind bubbles
and calculated the resulting X-ray luminosity.
The results differ from the simple model of
\citet{immler}.
We find that \citet{immler} assumptions overestimate the forward shock velocity, 
the cooling function and the luminosity of the shocked CSM region, while the
temperature of the forward shock region and the ratio $L_{rev}/L_{for}$
region are underestimated.
	\item The expected total luminosity of our numerical models is almost equal to or even higher than that of \citet{immler} analytical predictions, depending on the SN Ia 
explosion model used for our simulations. In particular, we note the 
significantly larger X-ray luminosity predicted for the W7 explosion model.
	\item We confirm the X-ray upper limits of the mass-loss rate of the donor star in a symbiotic progenitor system, placed by \citet{immler}, and in case of the W7 explosion model, stricter upper limits should be imposed.
	\item We simulated the mass outflows expected from SyRN systems in which the CSM is shaped by the interaction of a slow stellar wind with periodical nova explosions. The resulted CSM  consists of various nova shells with low density environments in between them.
	\item  We simulated the interaction of SNe Ia with  this wind-nova CSM. We find that the total SN X-ray luminosity is at least one order of magnitude less than that of a SN evolving in a wind bubble. We conclude that SNe Ia originating from SyRNe systems are not detectable in the X-rays during the early SN phase.  
	\item Our SyRN models reveal similar characteristics with  the CSM  of  SNe Ia which show time variable brightness (SN 2002ic) or time variable line abortion features (SN 2006X, 2007le, PTF 2011kx ).   
	
\end{enumerate}

\section*{Acknowledgments}
We are grateful to Carles Badenes and Eduardo Bravo for providing
us with the DDTc Type Ia model and Rony Keppens for providing
us with the AMRVAC code. We thank Sander Walg and K.M. Schure for their helpful discussions that improved the manuscript. Finally, it is a pleasure to thank the Lorentz Center Leiden for hosting a workshop on Type Ia progenitor systems in September 2013. This paper profited from the many discussion that took place there.

\label{lastpage}

\bibliographystyle{mn2e}
\bibliography{manuscript}

\end{document}